\documentclass[showpacs,twocolumn,pra]{revtex4}
\usepackage{amsmath}
\usepackage{amsfonts}
\usepackage{amssymb}
\usepackage{graphicx}
\usepackage[colorlinks=true,dvipdfm]{hyperref}

\begin{document}

\title{Experimental demonstration of deterministic one-way quantum computing
on a NMR quantum computer}
\author{Chenyong Ju}
\author{Jing Zhu}
\author{Xinhua Peng}
\email{xhpeng@ustc.edu.cn}
\author{Bo Chong}
\author{Xianyi Zhou}
\author{Jiangfeng Du}
\email{djf@ustc.edu.cn}
\affiliation{Heifei National Laboratory for Physical Sciences at Microscale and
Department of Modern Physics, University of Science and Technology of China,
230026 Hefei, People's Republic of China}

\begin{abstract}
One-way quantum computing is an important and novel approach to quantum
computation. By exploiting the existing particle-particle interactions, we
report the first experimental realization of the complete process of
deterministic one-way quantum Deutsch-Josza algorithm in NMR, including
graph state preparation, single-qubit measurements and feed-forward
corrections. The findings in our experiment may shed light on the future
scalable one-way quantum computation.
\end{abstract}

\maketitle

\section{Introduction}

The one-way quantum computing (QC) \cite{onewayPRL,onewayPRA} is a recently
proposed approach to quantum computation \cite{Chuang}. Being entirely
different to the traditional quantum circuit model \cite{Chuang}, it invokes
only single-qubit measurements with appropriate feed-forwards to accomplish
the computation, provided a highly entangled state - the cluster state \cite%
{clusterstate} or some other special shaped graph states \cite{othergraph} -
is given in advance. These entangled states serve as the universal resource
of the one-way QC. The one-way QC is not only important as a novel quantum
computing model, it also helps people to further understand the quantum
entanglement and measurement since these two fundamental physcial concepts
are particularly highlighted in this model.

The one-way QC has so far attracted much attention of the physical
community. Besides various theoretical researches, the existing experiments
mainly focused on the generation and characterization of a few-qubit graph
states \cite{eworkn1,ework1,ework2}, the demonstration of one- and two-qubit
gates \cite{eworkn1,eworkn2,eworkprl1,eworkprl2a}, and the realization of
two-qubit quantum algorithms \cite{eworkn1,eworkn2,eworkprl1,eworkdj}. Up to
now all the experiments of one-way QC were performed in linear optics. Owing
to the lack of interaction between photons, the cluster states were
generated probabilistically, and the success rate decreased exponentially
with the number of photons.

In this paper, by exploiting the existing particle-particle interactions, we
realized the deterministic one-way QC in NMR, including the graph state
generation, single-qubit measurement, and feed-forward. Our experiment
consists of two parts, the deterministic generation of a star-like four
qubit graph state and the implementation of a two-qubit Deutsch-Josza (DJ)
algorithm.

\section{Theory}

We consider the simplest version of the DJ \cite{DJ} algorithm, which
examines whether an unknown one-bit to one-bit function $f$ is constant or
balanced. There are four possible such functions as described in Fig. \ref%
{djoracle}(b). Classically one needs to call the function twice to check
both the outputs $f(0)$ and $f(1)$, while in DJ algorithm only one function
call is needed. The process of the algorithm is illustrated in Fig. \ref%
{djoracle}. The "function call" is implemented in the oracle, which applies
the following unitary operations: first a $\sigma _{z}$ applied on the
target qubit (denoted by $t$), then the two-qubit operation $|x\rangle
_{c}|y\rangle _{t}\rightarrow |x\rangle _{c}|y\oplus f_{j}(x)\rangle _{t}$ ($%
x,y\in \{0,1\}$) corresponding to the specific function $f_{j}$ ($j=1,2,3,4$%
). The result of the algorithm is read on the control qubit (denoted by $c$)
by measuring it with $\sigma _{x}$ (the Pauli operator). If the outcome is $%
|+\rangle $ then the function is definitely to be constant, otherwise it is
balanced.

\begin{figure}[tbp]
\begin{center}
\includegraphics[ width=0.48\textwidth ]{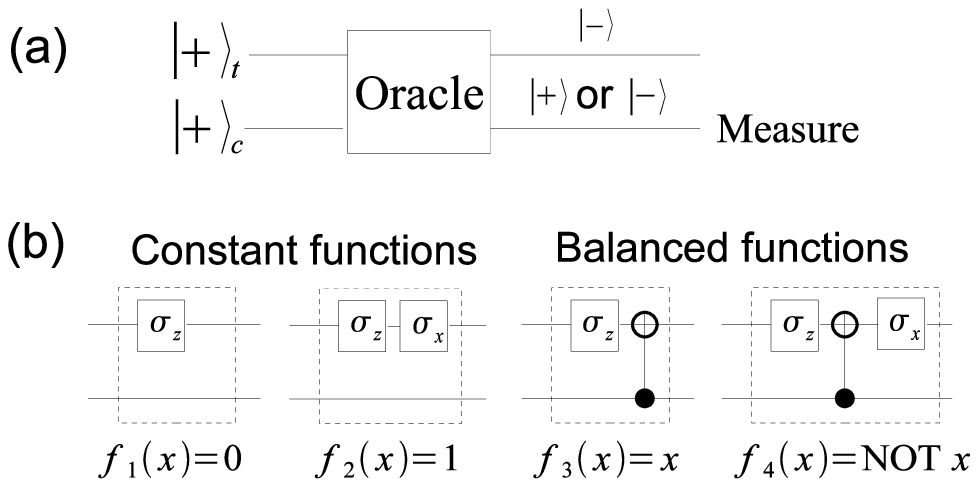}
\end{center}
\caption{(a) Schematic illustration of the Deutsch-Josza algorithm. (b) The
quantum networks for all possible oracles. The inter-line symbol in the
network denotes the controlled-NOT gate $| 0\rangle \langle 0| ^{(c)}\otimes
\mathbb{I}^{(t)}+| 1\rangle \langle 1| ^{(c)}\otimes \protect\sigma %
_{x}^{(t)}$, where the superscripts denote the qubits and $\mathbb{I}$ is
identity.}
\label{djoracle}
\end{figure}

\begin{figure}[tbp]
\begin{center}
\includegraphics[ width=0.45\textwidth ]{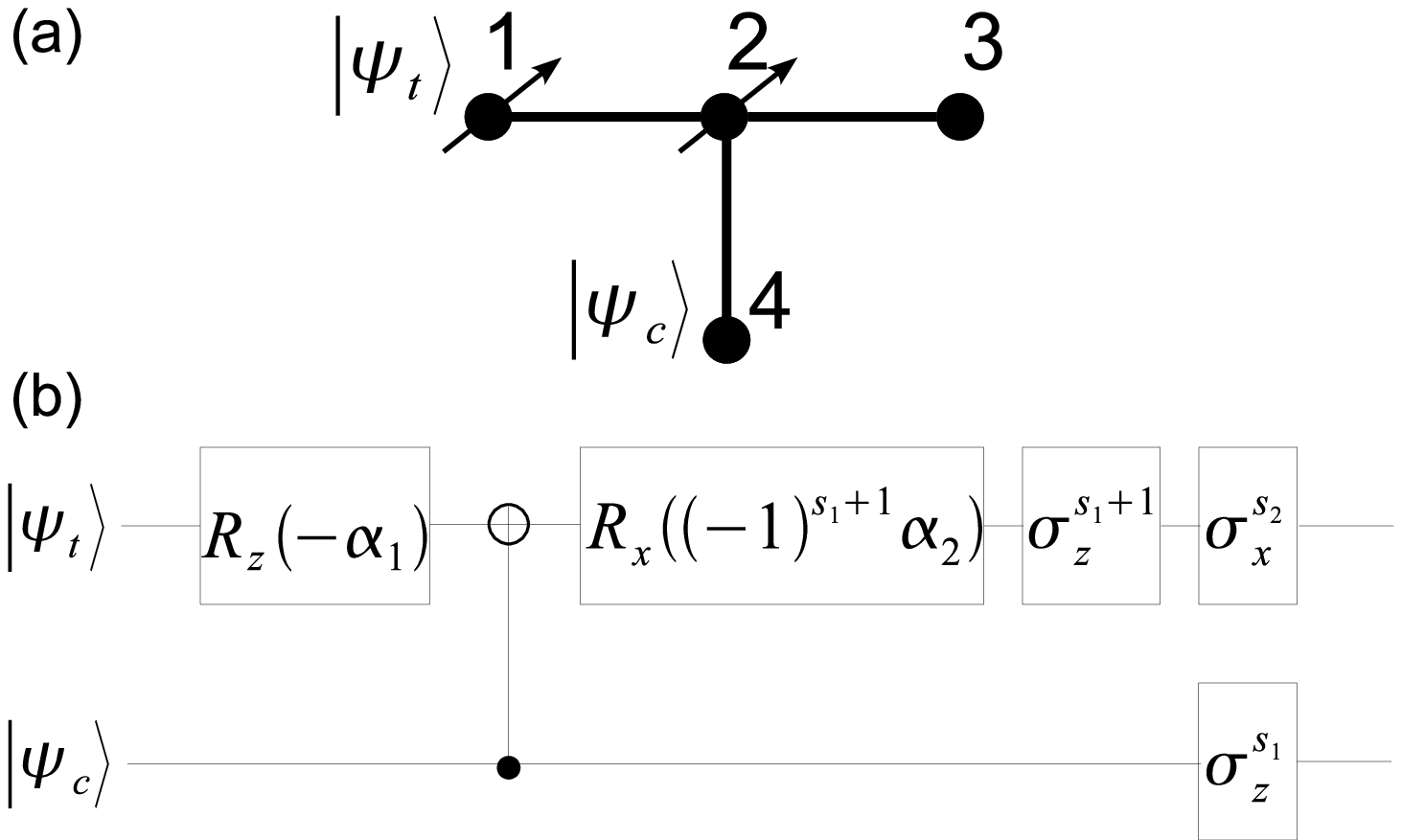}
\end{center}
\caption{(a) The star-like 4-qubit graph. The vertices represent the four
physical qubits, and each edge between them indicates a controlled-phase
gate applied on the connected qubits when to produce the entangled state.
The arrows marked on vertices 1 and 2 denote the measurements. For the
detailed description of the initial state, entangling, and measurements, see
the text. (b) The general logical quantum circuit realized on the above
graph with arbitrary logical input states and arbitrary measurement bases. $%
R_{n}(\protect\alpha )=e^{-i\protect\alpha \protect\sigma _{n}/2}$ $(n=x,z)$
}
\label{graph}
\end{figure}

To implement the DJ algorithm one should be able to construct all the
possible four configurations (Fig. \ref{djoracle}(b)). We find it is
sufficient to do these on the star-like 4-qubit graph state with appropriate
single-qubit measurement sequences and corresponding feed-forwards. We start
with the introduction of the general logical quantum circuit that can be
realized on the star-like 4-qubit graph (Fig. \ref{graph}(a)), with
arbitrary logical input states and arbitrary single-qubit measurement bases
in the $x$-$y$ plane. For this purpose, as in Ref. \cite{onewayPRL}, we
first prepare the four physical qubits in the graph into the initial state $%
|\psi _{t}\rangle _{1}\otimes |+\rangle _{2}\otimes |+\rangle _{3}\otimes
|\psi _{c}\rangle _{4}$, where the two arbitrary logical input states $|\psi
_{t}\rangle $ and $|\psi _{c}\rangle $ of the two logical qubits \cite%
{eworkn1} (a target qubit and a control qubit) are initially encoded on the
physical qubits 1 and 4. Then all physical qubits are entangled by the
entangling operator%
\begin{equation}
S=S^{(12)}S^{(23)}S^{(42)},  \label{etangling}
\end{equation}%
where $S^{(jk)}=|0\rangle \langle 0|^{(j)}\otimes \sigma
_{z}^{(k)}+|1\rangle \langle 1|^{(j)}\otimes \mathbb{I}^{(k)}$ is a
controlled-phase gate applied on the physical qubits $j$ and $k$ \cite%
{cphase}. The logical information is now delocalized. It is then manipulated
by the measurements carried on the physical qubits 1 and 2 with the bases $%
B_{1}(\alpha _{1})=\{|\alpha _{1+}\rangle _{1},|\alpha _{1-}\rangle _{1}\}$
and $B_{2}(\alpha _{2})=\{|\alpha _{2+}\rangle _{2},|\alpha _{2-}\rangle
_{2}\}$ respectively, where $|\alpha _{\pm }\rangle =\frac{1}{\sqrt{2}}%
(|0\rangle \pm e^{i\alpha }|1\rangle )$ and $\alpha _{1},\alpha _{2}$ are
arbitrary angles. After the measurements the logical target qubit is
transferred to the physical qubit 3, while the logical control qubit is
still encoded on the physical qubit 4. The effective logical quantum circuit
performed on the two logical qubits for the above process is shown in Fig. %
\ref{graph}(b). The $s_{1}$ and $s_{2}$ in the circuit denote the outcome of
the two measurements, where $s_{j}=0(1)$ corresponds to the output state $%
|\alpha _{j+}\rangle (|\alpha _{j-}\rangle )$ ($j=1,2$). The presence of the
$s_{1}$ and $s_{2}$ represents the randomness introduced by the single-qubit
projective measurements. In one-way QC appropriate feed-forwards \cite%
{onewayPRL} compensate for this randomness to restore the determinacy of the
computing.

The design of the resource entangled state and the single-qubit measurement
sequences for implementing the DJ algorithm is done by comparing the general
logical circuit in Fig. \ref{graph}(b) with the networks in Fig. \ref%
{djoracle}(b). Since the input state in the DJ algorithm is $| +\rangle
\otimes | +\rangle $, we set $| \psi _{t}\rangle =| \psi _{c}\rangle =|
+\rangle $. The resulted entangled state, $| \psi _{\text{G}}\rangle = (|
+\rangle _{1}| 0\rangle _{2}| -\rangle _{3}| +\rangle _{4}+| -\rangle _{1}|
1\rangle _{2}| +\rangle _{3}| -\rangle _{4})/\sqrt{2}$, which is produced by
performing the entangling operator $S$ on the initial state $| +\rangle
_{1}\otimes | +\rangle _{2}\otimes | +\rangle _{3}\otimes | +\rangle _{4}$,
is exactly the 4-qubit graph state which corresponds to the star-like
4-qubit graph \cite{onewayPRA} (it is also equivalent to the 4-qubit GHZ
state under local unitary operations). The measurement bases and the
corresponding feed-forward operations which are chosen to reproduce the
networks of all possible oracles are summarized in the Table \ref{table}.
Note in the design of the measurements which correspond to the constant
functions $f_{1}$ and $f_{2}$, the fact is used that a CNOT gate is
equivalent to the identity operation when it acts on the state $| +\rangle
\otimes | +\rangle $. The final result of the algorithm is read by measuring
the physical qubit 4. If its state (after accounting for the feed-forward
operation) is $| +\rangle (| -\rangle )$ then the function is
constant(blanced).

\begin{table}[tb]
\caption{Measurement bases and feed-forward operations for implementing the
DJ algorithm. $\text{FF}^{(3)}$ and $\text{FF}^{(4)}$ represent the
feed-forward operations on physical qubits 3 and 4.}
\label{table}
\begin{tabular*}{0.48\textwidth}{@{\extracolsep{\fill}}cccc}
\hline\hline
& Measurement bases & $\text{FF}^{(3)}$ & $\text{FF}^{(4)}$ \\ \hline
$f_{1}$ & $B_{1}(0),B_{2}(0)$ & $\sigma_{z}^{s_{1}}\sigma_{x}^{s_{2}+1}$ & $%
\sigma_{z}^{s_{1}}$ \\
$f_{2}$ & $B_{1}(0),B_{2}(0)$ & $\sigma_{z}^{s_{1}}\sigma_{x}^{s_{2}}$ & $%
\sigma_{z}^{s_{1}}$ \\
$f_{3}$ & $B_{1}(\pi),B_{2}(0)$ & $\sigma_{z}^{s_{1}+1}\sigma_{x}^{s_{2}}$ &
$\sigma_{z}^{s_{1}}$ \\
$f_{4}$ & $B_{1}(\pi),B_{2}(0)$ & $\sigma_{z}^{s_{1}+1}\sigma_{x}^{s_{2}+1}$
& $\sigma_{z}^{s_{1}}$ \\ \hline\hline
\end{tabular*}
\end{table}

\section{Experiment}

\subsection{Graph state generation}

\begin{figure}[tbp]
\begin{center}
\includegraphics[ width=0.45\textwidth ]{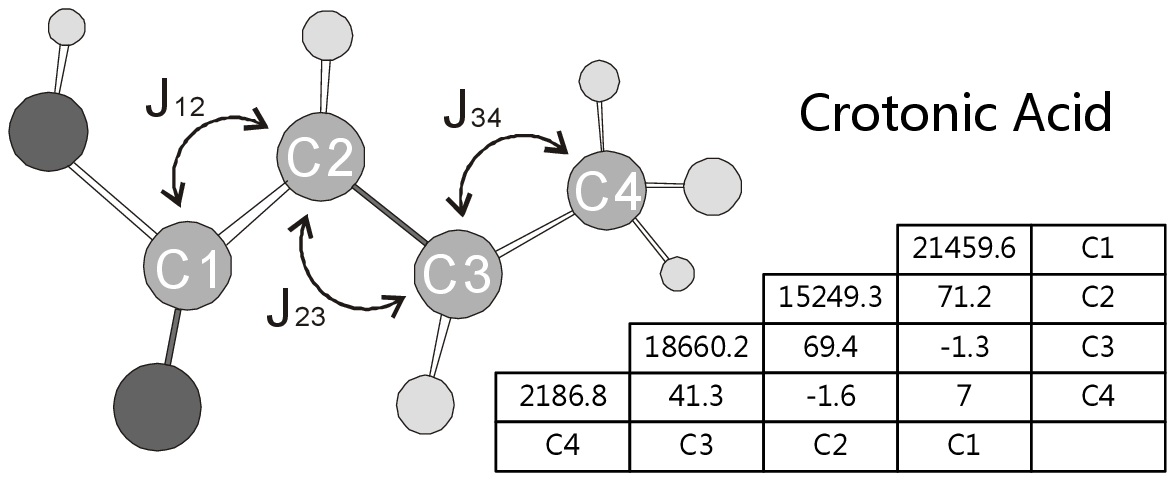}
\end{center}
\caption{Molecular structure of crotonic acid with a table of the chemical
shifts (on the diagonal) and $J$-coupling constants (below the diagonal).
The parameters are given in unit Hz.}
\label{crotacid}
\end{figure}

We used the spins of the four $^{13}$C nuclei in the crotonic acid dissolved
in $\text{D}_{2}\text{O}$ (see Fig. \ref{crotacid}) as the four physical
qubits, which provide the reduced hamiltonian $H_{0}=\sum_{j=1}^{4}\omega
_{j}I_{z}^{(j)}+2\pi \sum_{j<k}^{4}J_{jk}I_{z}^{(j)}I_{z}^{(k)}$ with the
Larmor angular frequencies $\omega _{j}$ and $J$-coupling strengths $J_{jk}$%
. On a Bruker UltraShield 500 spectrometer at room temperature, the measured
parameters are listed in Fig. \ref{crotacid} and their relaxation times are
obtained as $T_{1}(C1)=12.37s$, $T_{1}(C2)=4.89s$, $T_{1}(C3)=4.13s$, $%
T_{1}(C4)=4.96s$, $T_{2}(C1)=376.2ms$, $T_{2}(C2)=506.7ms$, $%
T_{2}(C3)=566.5ms$, $T_{2}(C4)=544.5ms$. In experiments, we labelled the
nuclei spins C2, C4, C3, C1 as the physic qubits 1, 2, 3, 4 in Fig. 2.

\begin{figure}[tbp]
\begin{center}
\includegraphics[ width=0.95\columnwidth]{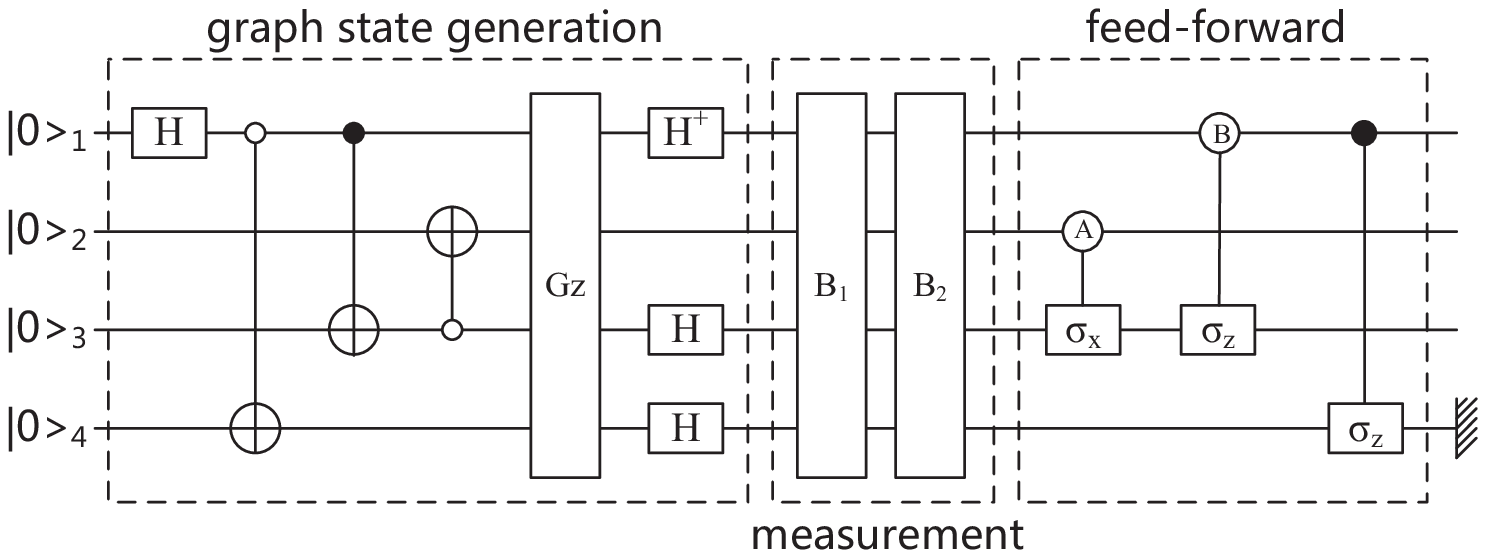}
\end{center}
\caption{Network for implementing the one-way DJ algorithm, consisting of
the main three stages: (i) graph state generation, (ii) single-qubit
measurements and (iii) feed-forward correction. Here $\text{H}=e^{-i \frac{%
\protect\pi}{4}\protect\sigma_y}$ is the pseudo-Hadamard gate with its
conjugate ${H^+}$, and $\text{G}_z$ is the pulsed magnetic filed gradient
along \textit{z} axis. In the feed-forward process, the controlled gates are
performed by the conditionality on the controlling spin being in the $\vert
0 \rangle$ or $\vert 1 \rangle$ state, which are determined by the values of
(A, B): (A, B) = (0, 1), (1, 1), (1, 0) and (0, 0) correspond to the
functions $f_1$ to $f_4$.}
\label{network}
\end{figure}


We first initialized the NMR ensemble to a standard pseudopure state with
deviation $|0000\rangle \langle 0000|$ from the thermal equilibrium state
using the spatial average technique \cite{spatial}. Then a 4-qubit GHZ state
$|\psi \rangle _{GHZ}=(|0110\rangle +|1001\rangle )/\sqrt{2}$ can be created
from $|0000\rangle $ by the network shown in Fig. \ref{network}. Here, the
controlled-not gate $\text{CNOT}^{(jk)}$ could be further decomposed into
\begin{equation}
R_{-z}^{(k)}(\frac{\pi }{2})-R_{-z}^{(j)}(\frac{\pi }{2})-R_{-x}^{(k)}(\frac{%
\pi }{2})-e^{-i\frac{\pi }{4}\sigma _{z}^{(j)}\sigma _{z}^{(k)}}-R_{y}^{(k)}(%
\frac{\pi }{2}),
\end{equation}%
where $R_{y}^{(k)}(\frac{\pi }{2})=e^{-i(\pi /4)\sigma _{y}^{(k)}}$ denotes
a $\frac{\pi }{2}$-rotation of the qubit $k$ around the $\hat{y}$ axis, and
so forth. The $J$-coupling gate $e^{-i\frac{\pi }{4}\sigma _{z}^{(j)}\sigma
_{z}^{(k)}}$ can be realized by the free evolution of the spin system under $%
H_{0}$ with appropriate refocusing pulses. Those $\hat{z}$ rotations in CNOT
gates can be omitted due to the initial state $|0000\rangle $\cite{nmrreview}%
. Such a GHZ state has the total zero spin quantum number that facilitates
us to clean-up the state further by a pulsed magnetic field gradient \cite%
{ernst}. Finally, the graph state $|\psi _{\text{G}}\rangle $ is generated
from the GHZ state by the local operations $R_{-y}^{(1)}(\frac{\pi }{2}%
)R_{-y}^{(3)}(\frac{\pi }{2})R_{-y}^{(4)}(\frac{\pi }{2})$.

In experiments, all the single-qubit operations are realized by sequences of
strongly modulating NMR pulses created by GRAPE algorithm \cite{grape}. We
maximized the gate fidelity of the simulated propagator to the ideal gate,
taking into account of radio frequency (RF) inhomogeneity. Theoretically the
gate fidelities we calculated for every pulse are greater than 0.99, and all
the pulse lengths are 600 $\mu s$.

\begin{figure}[tbp]
\begin{center}
\includegraphics[ width=0.48\columnwidth ]{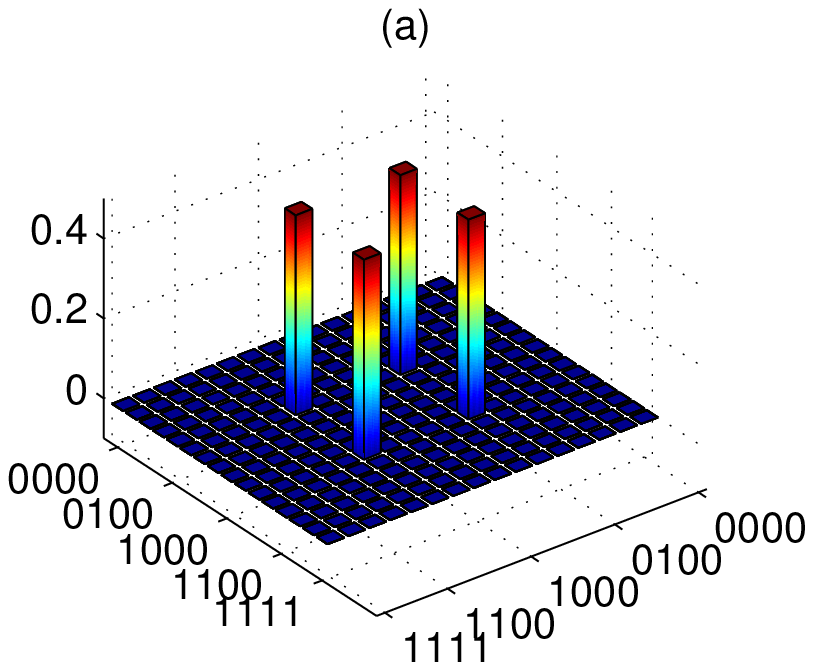}
\includegraphics[
width=0.48\columnwidth ]{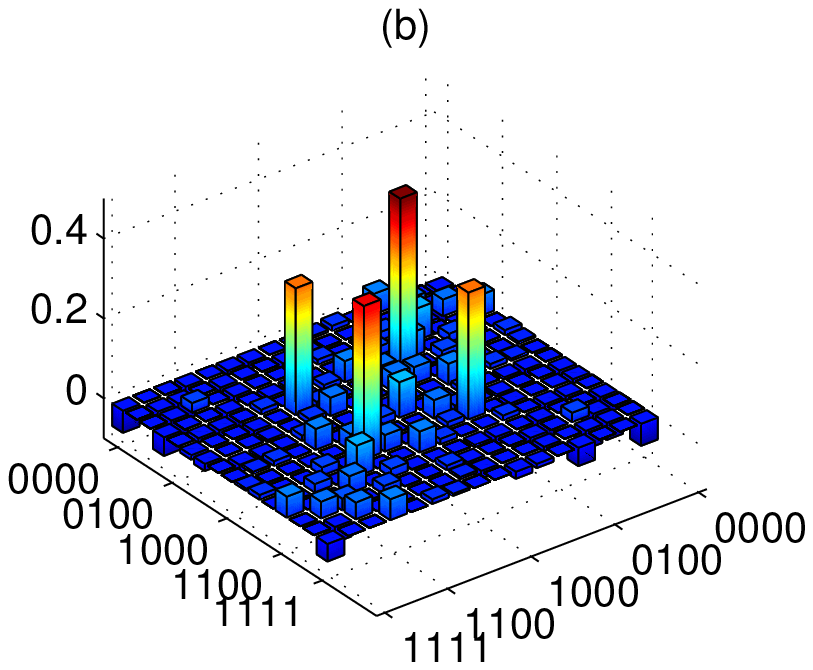}
\end{center}
\caption{(color online) The ideal (a) and measured (b) density matrices of
the four-qubit GHZ state $|\protect\psi\rangle_{GHZ}=(|0101\rangle+|1010%
\rangle)/\protect\sqrt{2}$. The rows and columns are numbered with the computational basis states. }
\label{exgraphstate}
\end{figure}

In order to access the quality of the graph state we prepared, we employed
the state tomography process \cite{tomo} to obtain the reconstructed density
matrix $\rho _{\text{exp}}$. Since only single-quantum coherences can be
directly observed, the tomography involves 19 repetitions of the experiment,
each with a different readout pulse sequence \cite{aaa} by
taking consideration of some unresolved J couplings. The real part of the reconstructed density matrix is shown in
Fig. \ref{exgraphstate} along with the theoretical expectation. The imaginary part is small comparable to zero in
the theoretical expectation. To
quantify how close the prepared state $\rho _{\text{exp}}$ to the
theoretical one $\rho _{id}$, we used the measure of the \textit{attenuated
correlation} defined by $c(\rho _{exp})=Tr(\rho _{id}\rho _{exp})/\sqrt{%
Tr(\rho _{id}^{2})}$, which takes into account both systematic errors and
the net loss of magnetization due to random errors and decoherence. The
value of the correlation for the tomographic readout of the GHZ state $|\psi
\rangle _{GHZ}$ is $c(\rho _{GHZ}^{exp})=0.73$, compared to the simulated
value 0.80 where we adopted the decoherence model \cite{corr} to take the
loss of magnetization due to spin relaxation. The experimental time for the
preparation is around 85 $ms$. To remove the effect of the net loss of
magnetization due to random errors and spin relaxation, we also calculated
the fidelity defined \cite{corr2} by $F_{GHZ}=Tr(\rho _{id}\rho _{exp})/%
\sqrt{Tr(\rho _{id}^{2})Tr(\rho _{exp}^{2}))}$, which yields the value of
0.88. Due to the imperfection of GRAPE pulses calculated and the effect of
strongly-coupling, the total theoretical fidelity is around 0.92. Moreover,
the lower experimental fidelity of 0.88 is mainly caused by other
uncertainties (e.g., the imperfection of the static magnetic field) in the
experiments. The correlation of the graph state $|\psi _{G}\rangle $ is
similar to the state $|\psi _{GHZ}\rangle $, because it was created only by
local operations.

Moreover, we confirmed the generation of the four-partite
pseudo-entanglement by obtaining a negative expectation value of the
entanglement witness \cite{witness} $W=I/2-| \psi _{\text{GHZ}}\rangle
\langle \psi _{\text{GHZ}}| $ ($I$ denotes identity operator) from the
tomographic readout: Tr$(W\rho _{\text{exp}})=-0.175 $.

\subsection{Implementation of DJ algorithm}

The required single-qubit projective measurements in the one-way QC are
absent in current NMR quantum computing technologies. But instead we can use
the pulsed magnetic field gradients to mimic them \cite{measure}, whose
effects are equivalent to that by applying these projective measurements on
every member of the ensemble. For example, one could use the operations%
\begin{gather}
P_{z}^{(1)}:G_{1}-R_{y}^{(2)}(\pi )-R_{y}^{(4)}(\pi )-G_{1}-R_{y}^{(3)}(\pi
)-R_{y}^{(4)}(\pi )  \notag \\
-G_{1}-R_{-y}^{(2)}(\pi )-R_{-y}^{(4)}(\pi )-G_{1}-R_{-y}^{(3)}(\pi
)-R_{-y}^{(4)}(\pi )
\end{gather}%
to mimic the $\sigma _{z}^{(1)}$ measurement on the physical qubit 1, where $%
G_{1}$ is the pulsed magnetic field gradient along $\hat{z}$ with a period
of $\tau _{1}/4$. These operations dephase all the coherences which are
associated with the transitions of qubit 1 in the density matrix $\rho
=|\Phi \rangle \langle \Phi |$ (suppose $|\Phi \rangle =|0\rangle _{1}|\phi
_{0}\rangle +|1\rangle _{1}|\phi _{1}\rangle $, where $|\phi _{0(1)}\rangle $
is the state of the other three qubits), resulting in an ensemble-average
density matrix $|0\rangle _{1}\langle 0||\phi _{0}\rangle \langle \phi
_{0}|+|1\rangle _{1}\langle 1||\phi _{1}\rangle \langle \phi _{1}|$, which
is exactly the same as the outcome by performing projective measurement $%
\sigma _{z}^{(1)}$ on every member of the ensemble. To mimic the measurement
$\sigma _{x}^{(1)}$, one just first rotate the qubit 1 to the $\hat{z}$ axis
before $P_{z}^{(1)}$. In our experiment we use the sequences $R_{-y}^{(1)}(%
\frac{\pi }{2})-P_{z}^{(1)}$ and $R_{y}^{(1)}(\frac{\pi }{2})-P_{z}^{(1)}$
to mimic the measurements with the bases $B_{1}(0)$ and $B_{1}(\pi )$
respectively. Therefore the different outcomes are labeled in the different
subspaces of qubit 1 denoted by $|0\rangle _{1}\langle 0|$(i.e., $s_{1}=0$)
and $|1\rangle _{1}\langle 1|$ (i.e., $s_{1}=1$). The measurement on qubit 2
with basis $B_{1}(0)$ is mimicked by a similar sequence $R_{-y}^{(2)}(\frac{%
\pi }{2})-P_{z}^{(2)}$, where%
\begin{gather}
P_{z}^{(2)}:R_{y}^{(3)}(\pi )-R_{y}^{(4)}(\pi )-G_{2}-R_{y}^{(1)}(\pi
)-R_{y}^{(4)}(\pi )-G_{2}  \notag \\
-R_{-y}^{(3)}(\pi )-R_{-y}^{(4)}(\pi )-G_{2}-R_{-y}^{(1)}(\pi
)-R_{-y}^{(4)}(\pi )-G_{2},
\end{gather}%
where $G_{2}$ is the pulsed magnetic field gradient with a period of $\tau
_{2}/4$.

\begin{figure}[tbp]
\begin{center}
\includegraphics[ width=0.9\columnwidth ]{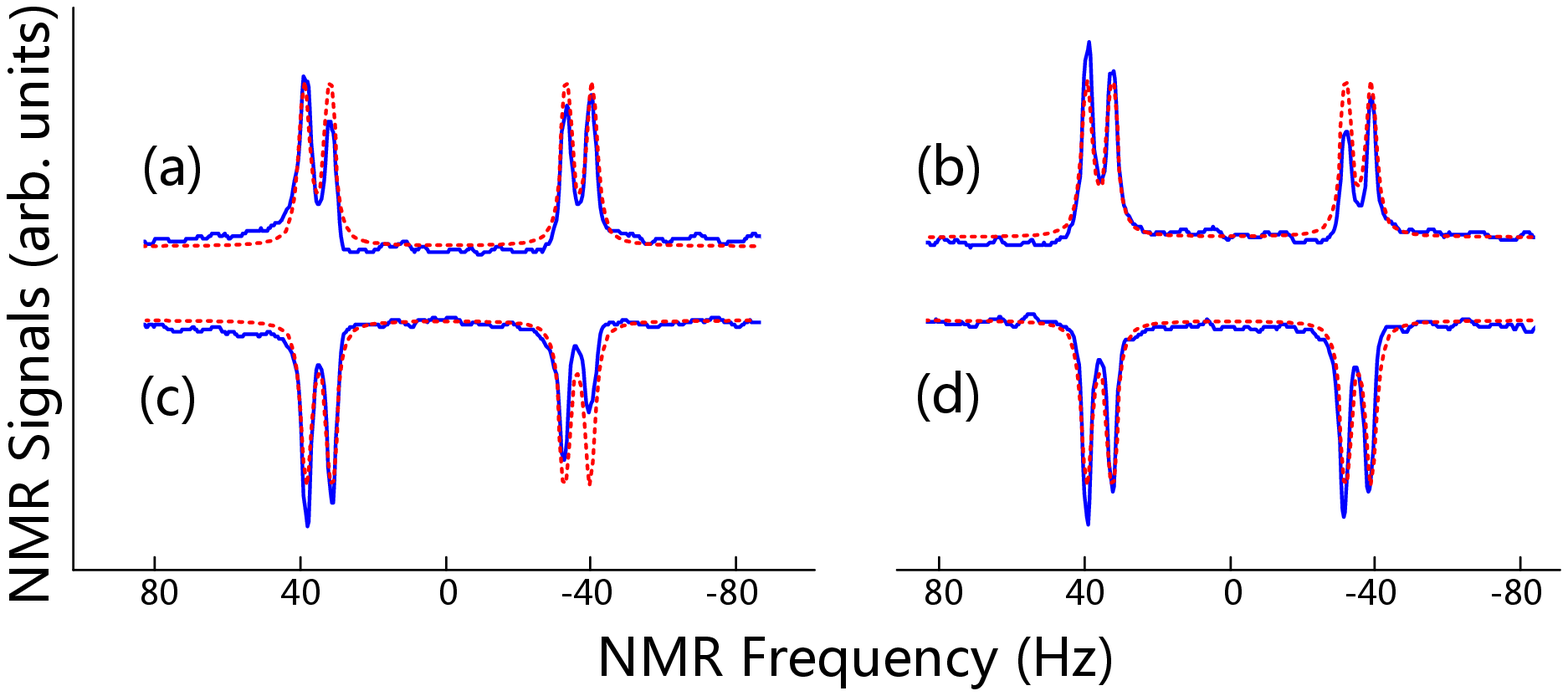}
\end{center}
\caption{(color online) The experimental (line) and simulated (dotted line)
spectra of qubit 4 for one-way DJ algorithm: (a)-(b) for the constant
functions $f_1$ and $f_2$, (c)-(d) for the balanced functions $f_3$ and $f_4$.
}
\label{finalresult}
\end{figure}

After the measurements, the computational outcomes have been stored in the
different subspaces depending on the measurement results $s_{1}$ and $s_{2}$%
. However, to obtain a deterministic and correct computation, the
feed-forward operations should be carried out conditionally on the
measurement results $s_{1}$ and $s_{2}$. We also experimentally realized the
feed-forward operations by the conditional unitary operations illustrated in
Fig. \ref{network}. After this, the judgement of whether the function $f$ is
constant or balanced is determined by measuring the state of qubit 4: if the
state of qubits is in $|+\rangle $, the function $f$ is constant;
conversely, if the state of qubits is in $|-\rangle $, the function $f$ is
balanced. As a result of $|\pm \rangle \langle \pm |=\frac{1}{2}I\pm \sigma
_{x}$, the $|+\rangle /|-\rangle $ state will give a positive/negative NMR
signal if we first reference the spectrum of the thermal equilibrium after a
$[\frac{\pi }{2}]_{y}$ pulse as a positive one. The results are shown in
Fig. \ref{finalresult} for four cases $f_{1}-f_{4}$. In (a) and (b), the
signal amplitude is positive, which indicates the state of qubit 4 is in $%
|+\rangle $ and the function $f$ is constant. In (c) and (d), the signal is
inverted which indicates the state of qubit 4 is in $|-\rangle $ and the
function $f$ is balanced. Here the carbon signal-to-noise ratio
(Signal/Noise) of the spectra is about 44.4-48.3. The imperfection of the
experimental spectra compared to the simulated spectra is mainly caused by
the unideal input graph state, the imperfection of the static magnetic field and the GRAPE pulses. Alternatively, the feed-forward process can also be replaced by the
single-qubit measurement under a suitable basis \cite{onewayPRL}, which
makes that the one-way QC can be completed solely by single-qubit
measurements.

\section{Conclusion}

We realized the complete process of deterministic one-way QC using a
star-like 4-qubit graph state in NMR: the graph state was generated
deterministically by exploiting the existing particle-particle interactions
in our experiments; the single qubit measurements were mimicked by pulsed
magnetic field gradients; and the feed-forward corrections were realized by
conditional-unitary gates in our specific NMR system. Here NMR technique is
used as a test-bed for the demonstration of deterministic one-way QC,
however, it does not imply that any of the results obtained are specific to
NMR. The experience obtained from our experiments could be helpful to other
physical systems for future scalable deterministic one-way QC and deserve
further investigation. If the direct particle-particle interactions of the
quantum system (such as trapped ions and optical lattices) are provided, one
can prepare the cluster state deterministically, do single-qubit
measurements, and perform the feed-forward correction by unitary gates,
therefore realizing scalable one-way QC deterministically.

\begin{acknowledgments}
We thank V.Vedral for helpful discussions. This work was supported by the
National Natural Science Foundation of China, the CAS, and the National
Fundamental Research Program.
\end{acknowledgments}


\begin{thebibliography}{99}
\bibitem{cphase} Note that some papers use another controlled-phase gate $%
S=\left\vert 0\right\rangle \left\langle 0\right\vert \otimes \mathbb{I}%
+\left\vert 1\right\rangle \left\langle 1\right\vert \otimes \sigma _{z}$ as
the entangling operator. The produced graph states with these two entangling
operators are equivalent up to local unitary operations.
\end{thebibliography}
\end{document}